\documentclass[runningheads]{llncs}
\usepackage[T1]{fontenc}
\usepackage{multirow,multicol}

\usepackage{kantlipsum,arydshln}

\usepackage{amsmath}
\usepackage{tabularx}
\usepackage[table]{xcolor}
\usepackage{acronym}
\usepackage[skip=0pt]{caption}
\usepackage{pifont}
\usepackage{xspace}
\usepackage{subcaption}
\usepackage{makecell}
\usepackage{CJKutf8}
\usepackage[inline]{enumitem}
\usepackage{booktabs} 
\usepackage{url}
\usepackage{adjustbox}
\usepackage[colorlinks=true,
            linkcolor=blue,
            citecolor=blue,
            filecolor=blue,
            urlcolor=blue]{hyperref}

\usepackage{tcolorbox}

\renewcommand{\tablename}{Tab.}

\newcommand{\header}[1]{\vspace*{1mm}\noindent\textbf{#1}.}

\acrodef{BAC}{balanced accuracy}
\acrodef{CoT}{chain-of-thought}
\acrodef{IR}{information retrieval}
\acrodef{PLM}{pre-trained language model}
\acrodef{LLM}{large language model}
\acrodef{CIS}{conversational information seeking}
\acrodef{CNP}{clarification need prediction}
\acrodef{QPP}{query performance prediction}
\acrodef{RL}{reinforcement learning}

\newcommand{\our}{Zef-CNP\xspace}
\newcommand{\prompt}{TIQ-CoT\xspace}
\newcommand{\seq}{CoQu\xspace}
\newcommand{\data}{CoQu-CS\xspace}

\begin{document}

\title{Zero-Shot and Efficient Clarification Need Prediction in Conversational Search}
\titlerunning{Zero-Shot and Efficient Clarification Need Prediction}
%

\author{Lili Lu\inst{1}\orcidID{0009-0006-9547-1172} \and
Chuan Meng\inst{2}\orcidID{0000-0002-5070-2049} \and
Federico Ravenda\inst{1}\orcidID{0009-0004-1421-252X} \and
Mohammad Aliannejadi\inst{2}\orcidID{0000-0002-9447-4172} \and 
Fabio Crestani\inst{1}\orcidID{0000-0001-8672-0700}}

\authorrunning{Lu et al.}
%

\institute{Università della Svizzera italiana, Switzerland \\ \email{\{lili.lu,federico.ravenda,fabio.crestani\}@usi.ch} \and
University of Amsterdam, The Netherlands \\ 
\email{\{c.meng,m.aliannejadi\}@uva.nl}
}

\maketitle              
\begin{abstract}

\Ac{CNP} is a key task in conversational search, aiming to predict whether to ask a clarifying question or give an answer to the current user query.
However, current research on \ac{CNP} suffers from the issues of limited \ac{CNP} training data and low efficiency.
In this paper, we propose a zero-shot and efficient \ac{CNP} framework (\our), in which we first prompt \acp{LLM} in a zero-shot manner to generate two sets of synthetic queries: ambiguous and specific (unambiguous) queries. 
We then use the generated queries to train efficient \ac{CNP} models. 
\our eliminates the need for human-annotated clarification-need labels during training and avoids the use of \acp{LLM} with high query latency at query time.
To further improve the generation quality of synthetic queries, we devise a topic-, information-need-, and query-aware \ac{CoT} prompting strategy (\prompt).
Moreover,  we enhance \prompt with counterfactual query generation (\seq), which guides \acp{LLM} first to generate a specific/ambiguous query and then sequentially generate its corresponding ambiguous/specific query.
Experimental results show that \our achieves superior \ac{CNP} effectiveness and efficiency compared with zero- and few-shot \ac{LLM}-based \ac{CNP} predictors. %

\end{abstract}
\section{Introduction}

Recent years have witnessed considerable progress in mixed-initiative conversational search~\cite{zhang2024ask}, where the user and system can both take the initiative at different times in a conversation~\cite{radlinski2017theoretical}.
Since users often fail to express their information needs, asking clarifying questions is a crucial system-initiative action to address ambiguity in user queries~\cite{aliannejadi2024interactions,aliannejadi2019asking,deng2023prompting}.
Most existing work has focused on the selection or generation of clarifying questions~\cite{aliannejadi2019asking,sekulic2021towards,wang2023zero}, with relatively less attention given to determining when to ask them.
In this paper, we focus on the \acf{CNP} task, which is to predict whether to ask a clarifying question to the current user query~\cite{aliannejadi2021building,arabzadeh2022unsupervised,chen2024learning,deng2023learning,roitman2019study,wang2021controlling,wang2022simulating,xu2019asking}.
\ac{CNP} is a critical task because asking clarifying questions at an inappropriate time would hurt user experience~\cite{meng2023system}.

\header{Motivation}
Existing studies on \ac{CNP} have two main limitations:
\begin{enumerate}[label=(\roman*),nosep,leftmargin=*]
\item \textbf{Limited or no training data for \ac{CNP}}.
Manually annotating clarification-need labels is labor-intensive, which restricts the scale of \ac{CNP} training data.
For example, ClariQ, a widely-used \ac{CNP} dataset~\cite{aliannejadi2021building}, contains only 187 human-annotated clarification-need labels for training.
In some domain-specific conversational search scenarios, there is no training data for \ac{CNP}.

\item \textbf{Low efficiency}. Although existing work \cite{arabzadeh2022unsupervised} proposes an unsupervised method for \ac{CNP}, which analyzes retrieved documents for the current user query, this type of method is inefficient because it performs \ac{CNP} after retrieval.
Large language models (\acp{LLM}) have achieved promising results on various \ac{IR} tasks \cite{askari2024generative,meng2024query}.
One might argue that zero-shot prompting of \acp{LLM} has the potential to achieve promising results on \ac{CNP} \cite{deng2023prompting,kuhn2022clam,zhang2024clamber,zhang2024ask}.
However, \acp{LLM} with billions of parameters result in a substantial increase in computational overhead, making it hard to apply them in practice \cite{meng2024ranked}.
\end{enumerate}

\header{A new framework for \ac{CNP}}
In this paper, we aim to address these limitations by proposing a \textbf{z}ero-shot and \textbf{ef}ficient \ac{CNP} framework (\our), in which we first use zero-shot prompting with \acp{LLM} to generate synthetic specific queries (which do not require clarification) and ambiguous queries. 
These pseudo queries are then utilized to train efficient \ac{CNP} models, which do not rely on retrieval or \acp{LLM}. 
We only employ the trained efficient \ac{CNP} models during inference.
\our allows us to avoid using human-annotated clarification-need labels during training, and avoid using retrieval/\acp{LLM} introducing high latency at query time.

\header{Challenges}
However, \our presents new challenges.
Prompting \acp{LLM} to generate synthetic queries in a single attempt results in limited generation quality. 
Especially without the context of topics or information needs, it is challenging for humans to identify specific and ambiguous queries.

\header{Solution}
To address the challenge, we devise a \textbf{t}opic-, \textbf{i}nformation-need-, and \textbf{q}uery-aware \ac{CoT} prompting strategy (\prompt).
In \prompt, we instruct \acp{LLM} to generate a topic, an information need, step by step, before generating synthetic queries.
Furthermore, 
instead of generating specific and ambiguous queries separately, we enhance \prompt with \textbf{co}unterfactual \textbf{qu}ery generation (\seq), which guides \acp{LLM} to first generate a specific/ambiguous query and then sequentially generate its corresponding ambiguous/specific query, based on the same topic and information need.

\header{Experiments}
In \our, we prompt 2 \acp{LLM} (i.e., Llama-3.1-8B-Instruct, and GPT-4o-mini) for synthetic query generation.
With each \ac{LLM}, we generate 2,500 specific queries and 2,500 ambiguous ones.
We then fine-tune BERT~\cite{devlin2019bert} on synthetic specific and ambiguous queries for \ac{CNP}.
Finally, we evaluate the fine-tuned BERT on two \ac{CNP} datasets, ClariQ~\cite{aliannejadi2021building}, and AmbigNQ~\cite{min-etal-2020-ambigqa}. 
Experimental results show that \our (fully zero-shot) achieves superior \ac{CNP} effectiveness and efficiency compared with baselines directly using \acp{LLM} as \ac{CNP} predictors.
Our analysis indicates that our devised \prompt and \seq markedly improve data generation quality.

Our main contributions are as follows:
\begin{enumerate}
    \item We propose \our, a zero-shot and efficient \ac{CNP} framework, which does not rely on any human-annotated clarification-need labels. 
    \item We devise \prompt, a topic-, information-need- and query-aware \ac{CoT} prompting strategy, which instructs \acp{LLM} to first generate a topic and information need before generating queries. 
    Moreover, we improve \prompt with counterfactual query generation (\seq), which guides \acp{LLM} first to generate a specific/ambiguous query and then sequentially generate its corresponding ambiguous/specific query.
    \item Experimental results on two \ac{CNP} benchmark datasets show that \our achieves superior \ac{CNP} quality compared to baselines directly using \acp{LLM} as \ac{CNP} predictors.
    
    \item We release our generated synthetic data, called Counterfactual Queries for Clarification in Conversational Search (\data), to the public.
    Codes and data are available at~\url{ https://github.com/lulili0963/Zef-CNP}.
\end{enumerate}

\section{Related Work}
\label{sec:rw}

\header{Clarification need prediction}
\label{sec:rw:cnp}
Conversational search has attracted increasing attention~\cite{meng2023Performance,meng2020refnet,meng2021initiative,meng2020dukenet,zamani2023conversational}.
Mixed initiative is a key part of conversational search, which means that both the system and the user can take the initiative at any time during the conversation~\cite{radlinski2017theoretical}.
The system can take the initiative of performing various actions, such as asking for clarification~\cite{aliannejadi2019asking} or eliciting user preferences~\cite{kostric2024generating}.
In this paper, we focus on clarification.
The literature focuses on two main tasks in clarification: \acf{CNP} and clarifying question selection/generation~\cite{aliannejadi2019asking,erbacher2024paqa,sekulic2021towards,sekulic2024estimating}.
Our work focuses on \ac{CNP} in mixed-initiative conversational search, which is to predict whether to ask a clarifying question to the current user query in an information-seeking conversation.

Most studies~\cite{aliannejadi2021building,chen2024learning,kim2021deciding,meng2023system,xu2019asking} rely on training data to train an effective \ac{CNP} model.
For example, Aliannejadi et al.~\cite{aliannejadi2021building} and Chen et al.~\cite{chen2024learning} 
fine-tuned \acp{PLM} (e.g., BERT) and \acp{LLM} on human-annotated clarification-need labels, respectively.
However, manually annotating clarification-need labels is labor-intensive, restricting the scale of \ac{CNP} training data.
For example, Aliannejadi et al.~\cite{aliannejadi2021building} curated the ClariQ dataset, involving human annotators to have clarification-need labels, however, ClariQ only contains 187 annotated queries for training \ac{CNP}.
Methods trained on these limited datasets could become biased towards the datasets and lack generalizability~\cite{arabzadeh2022unsupervised}.

Some studies do not rely on human-annotated clarification-need labels.
For example, the two works of Wang and Ai~\cite{wang2021controlling,wang2022simulating} focused on training \ac{CNP} purely via reinforcement learning, where rewards are based on these human-designed simulators, however, the use of human-designed simulators limits the generalizability of \ac{CNP} methods.
Arabzadeh et al.~\cite{arabzadeh2022unsupervised} proposed an unsupervised method, which predicts \ac{CNP} by analyzing the retrieved documents for the current user query; this type of method is inefficient because it performs \ac{CNP} after retrieval.

Recently, some work~\cite{deng2023prompting,kuhn2022clam,zhang2024clamber,zhang2024ask} focused on prompting \acp{LLM} in a zero/few-shot manner for \ac{CNP}.
However, directly using \acp{LLM} as clarification-need predictors is highly inefficient and impractical due to their substantial computational overhead, as they have billions of parameters.
In particular, Deng et al.~\cite{deng2023prompting} and Zhang et al.~\cite{zhang2024clamber} used a \ac{CoT} prompting strategy for \ac{CNP}, i.e., they first prompted \acp{LLM} to generate descriptive reasoning before deciding whether clarification is necessary.
The additional tokens produced by \ac{CoT} further degrade the \ac{CNP} efficiency.

Our work differs as
\begin{enumerate*}[label=(\roman*)]
\item we aim at fully zero-shot \ac{CNP}, where we do not rely on any human-annotated clarification-need labels,
\item we focus on efficient \ac{CNP}, where we do not rely on retrieval or \acp{LLM} at query time.
\end{enumerate*}
Note that unlike the work from Deng et al. ~\cite{deng2022pacific}, Guo et al. ~\cite{guo2021abg} and Xu et al. ~\cite{xu2019asking}, focusing on \ac{CNP} scenarios where the ambiguity of a query depends on a given document or entities, we address a more general \ac{CNP} setting in conversational search, where \ac{CNP} models only have access to the current user query without additional contextual information.

\header{Counterfactual generation}
Leveraging LLMs for counterfactual generation has gained popularity and proven to be impactful in various IR and natural language processing (NLP) applications~\cite{abolghasemi2024cause,bhattacharjee2024towards,bhattacharjee2024zero,chen2022disco,dixit2022core}.
It refers to generating a query/utterance/question with a specific label that is the opposite to a given label.
For example, Abolghasemi et al.~\cite{abolghasemi2024cause} proposed using LLMs to generate counterfactual dialogues in which the simulated user is not satisfied with the system's response. They showed that this approach leads to a more balanced distribution of labels, resulting in higher satisfaction prediction performance in task-oriented dialogue systems. 
Calderon et al.~\cite{calderon2022docogen} 
generated counterfactual utterances for multi-label intent prediction. They produced domain-counterfactual samples similar to the original ones in every aspect but different in their domain.
Bhattacharjee et al.~\cite{bhattacharjee2024zero} introduced a zero-shot pipeline for generating high-quality counterfactuals, without any additional training or fine-tuning, indicating the effectiveness of the pipeline in a zero-shot manner.

Our work is aligned with the idea of counterfactual generation using LLMs, however, we take a different angle and focus on generating both specific queries and their counterfactual ambiguous ones in conversational search.

\header{Zero-shot synthetic data generation}
\label{sec:rw:syn}
Given the high cost of obtaining human-annotated data, zero-shot dataset generation has become a popular approach. 
It often contains 3 steps, which are the generation of synthetic data by \acp{PLM} or \acp{LLM}, the fine-tuning of \acp{PLM} or smaller models on the generated data, and the evaluation of fine-tuned models on various downstream tasks~\cite{gao2022zerogen,meng2022generating,ye2022zerogen,ye2022progen,zou2024fusegen}.
Meng et al.~\cite{meng2022generating} presented a method, called SuperGen, leveraging different \acp{PLM} for data generation and fine-tuning, and evaluating various classification tasks. 
However, the generative capabilities of \acp{PLM} are still limited, compared to \acp{LLM}.
Ye et al.~\cite{ye2022zerogen} performed an extremely similar pipeline, called ZEROGEN, and it relies on smaller models (e.g., DistilBERT~\cite{sanh2019distilbert}) for fine-tuning.

However, it differs as
\begin{enumerate*}[label=(\roman*)]
\item we focus on \ac{CNP} in conversational search, a domain where zero-shot dataset generation has not been explored, 
\item to improve synthetic query generation quality, we propose \prompt to ask \acp{LLM} to  generate topics and information needs before generating synthetic queries, and further introduce \seq to generate specific and ambiguous queries in a counterfactual way, and
\item beyond generating synthetic specific queries, we explore generating \textit{synthetic ambiguous queries}, which has not been well investigated yet.
\end{enumerate*}

\section{Task definition} 
Previous studies \cite{aliannejadi2021building,deng2023prompting} generally define \ac{CNP} as a binary classification task. 
Given the current user query $q$, a clarification-need predictor $f$ determines whether the query is ambiguous or not. 
Formally, $l^* = f(q)$, where $l^* \in \{0, 1\}$, 
$l^*$ is a predicted clarification-need label.
A label of $1$ indicates that the query $q$ is ambiguous and requires clarification, while a label of $0$ implies that the query is specific and does not need clarification.

\begin{figure}[t!]
    \centering
    \includegraphics[width=1.0\textwidth]{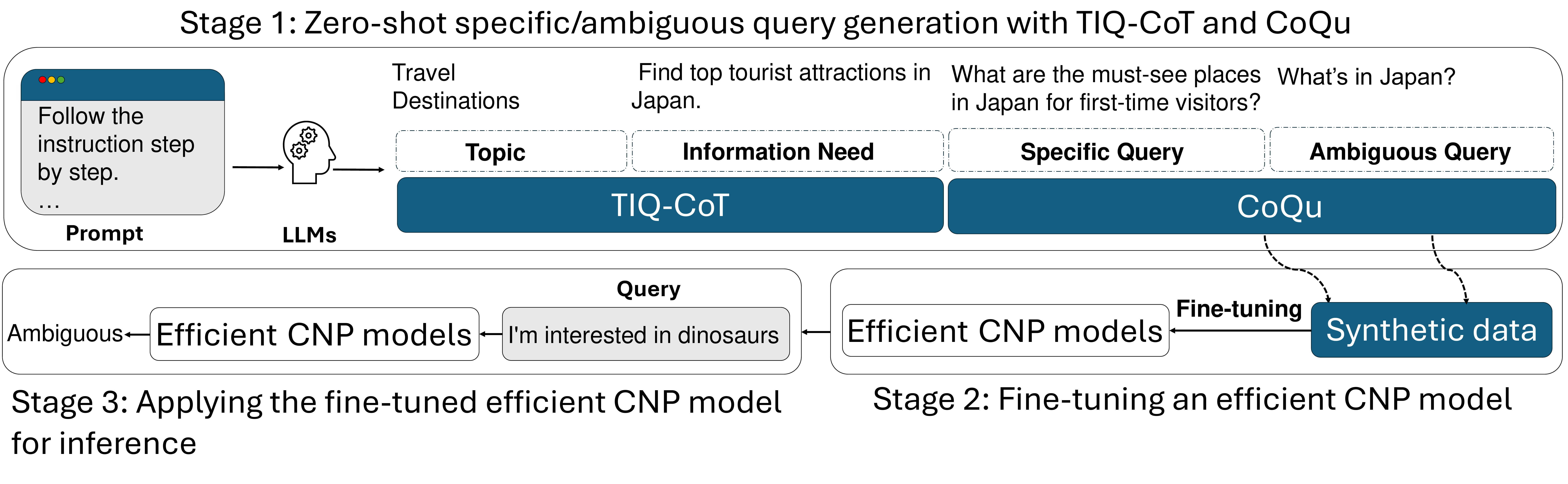} 
    \caption{Our proposed zero-shot and efficient \ac{CNP} framework (\our). Please be aware that we implement \prompt and \seq in the same prompt (see in Fig.~\ref{fig:prompt}). 
    }
    \label{fig:frame}
\end{figure}

\section{Methodology}
\label{sec:method}

We propose \our, a zero-shot and efficient \ac{CNP} framework.
As illustrated in Fig.~\ref{fig:frame}, it consists of three main stages:
\begin{enumerate*}[label=(\roman*)]
\item zero-shot specific/ambiguous query generation with \prompt and \seq,
\item fine-tuning efficient \ac{CNP} models, and
\item applying fine-tuned efficient \ac{CNP} models for inference.
\end{enumerate*}

\begin{figure}[tbp]
\centering
     \begin{tcolorbox}[notitle,boxrule=1pt,colback=gray!5,colframe=black, arc=2mm,width=\columnwidth]
\sf\textbf{Instruction}: Follow the instruction step by step. 1. Come up with 10 topics. 2. Based on the topics, generate 10 corresponding user information needs. 3. Based on the user information needs, generate 10 corresponding specific queries. 4. Based on the user information needs, generate 10 corresponding ambiguous queries. \\
The output format should be Topic: \{topic\}, User information need: \{user information need\}, Specific query: \{specific query\}, Ambiguous query: \{ambiguous query\}.
    \end{tcolorbox}
\caption{A topic-, information-need-, and query-aware \ac{CoT} prompting strategy (\prompt) enhanced with counterfactual query generation (\seq).}
\label{fig:prompt}
\end{figure}

\subsection{Zero-shot specific/ambiguous query generation with \prompt and \seq} 
This stage aims to generate $M$ synthetic training examples for \ac{CNP}, i.e., $\mathcal{D}=\{(\hat{q}_i,\hat{l}_i)\}^{M}_{i=1}$, where if $\hat{q}_i$ is a specific query (denoted as $sq$), then $\hat{l}_i = 0$; if $\hat{q}_i$ is an ambiguous query (denoted as $aq$), $\hat{l}_i = 1$.

We completely leverage \acp{LLM}' self-knowledge to generate synthetic specific/ambiguous queries in a purely zero-shot manner.
To accomplish effective specific/ambiguous query generation, we propose a \textit{topic-, information-need-, and query-aware \ac{CoT} prompting strategy} (\prompt), which is accompanied by our devised \textit{counterfactual query generation} (\seq).

\subsubsection{\prompt.}
Specifically, \prompt aims to first prompt (the input prompt denoted as $p$) an \ac{LLM} to generate a synthetic topic $t$, and a synthetic user information need $in$, before generating a specific ($sq$) or an ambiguous query ($aq$).
Formally:
\begin{equation}
t,in, sq/aq=\mathrm{LLM}(p).
\label{q1}
\end{equation}

\subsubsection{\seq.}
\seq further improves \prompt by instructing the \ac{LLM} to generate a specific query ($sq$), and then generating its corresponding ambiguous query ($aq$) in a counterfactual manner.
\seq can potentially help the \ac{LLM} more clearly differentiate between the two types of queries.
We formulate \prompt with \seq in the following:
\begin{equation}
t,in, sq, aq=\mathrm{LLM}(p).
\label{e2}
\end{equation}
In this way, we can collect a pair of specific and ambiguous queries ($sq$, $aq$) based on the same topic and information need.
We hypothesize that generating specific and ambiguous queries in a counterfactual manner helps \acp{LLM} better distinguish between the two types, to improve generation quality for both query types.
We also examine the generation order of specific and ambiguous queries in Tab.~\ref{tab:ablation}.

Specifically, we use the input prompt $p$ shown in Fig.~\ref{fig:prompt} to instruct an \ac{LLM} to generate ten pairs of specific and ambiguous queries per call.
After data generation, we conduct data filtering: we remove duplicates, empty values, and special characters.

\subsection{Fine-tuning an efficient \ac{CNP} model}
We use synthetic training data
$\mathcal{D}=\{(\hat{q}_i,\hat{l}_i)\}^{M}_{i=1}$ to train an efficient \ac{CNP} model $f$, which does not rely on \acp{LLM} or take retrieved information as inputs, ensuring high efficiency.
The learning objective is defined as:
\begin{equation}
\label{learning}
\begin{split}
\mathcal{L}(\theta)&=
-\frac{1}{M}\sum_{i=1}^{M}
\log P(\hat{l}_i \mid \hat{q}_i),
\end{split}
\end{equation}
where $\theta$ are all the parameters of $f$.

\subsection{Applying the fine-tuned efficient \ac{CNP} model for inference} 
After training the efficient \ac{CNP} model, we apply it to a test user query $q$ for inference. 
The model predicts whether the query requires clarification, denoted as $l^* = f(q)$, where $l^*$ is the predicted clarification-need label.

\section{Experiments}
\label{sec:method:experimental}

\header{Research questions} 
Our work is steered by the following research questions:
\begin{enumerate}[label=\textbf{RQ\arabic*}]

\item To what extent does \our improve the \textit{quality} and \textit{efficiency} of \ac{CNP}, compared to baselines that use \acp{LLM} as zero- or few-shot predictors for clarification needs? \label{RQ1} 
\item To what extent does the choice of \acp{LLM} for synthetic data generation in \our affect \ac{CNP} quality? \label{RQ2} 
\item To what extent do our devised \prompt and \seq improve \ac{CNP} quality? Does the generation order of specific and ambiguous queries in \seq affect \ac{CNP} quality?\label{RQ3} 
\item To what extent does the scale of synthetic specific/ambiguous queries generated by \our affect \ac{CNP} quality? \label{RQ4} 
\end{enumerate}

\begin{table}[t!]
\centering
\caption{Statistics of ClariQ, and AmbigNQ for evaluation.}
\label{statistics}
\begin{tabular}{lrrrr}
\hline
                         & ClariQ            &AmbigNQ          
                         \\ \hline

$\#$of ambiguous queries &262          &1,172          
\\\hline
$\#$of unambiguous queries &37         &830          
\\\hline
avg tokens of queries      &9.36       &12.75        
\\
\hline
\end{tabular}
\end{table}

\header{Datasets} 
Following \cite{aliannejadi2021building,arabzadeh2022unsupervised}, we use two datasets: ClariQ~\cite{aliannejadi2021building}, AmbigNQ~\cite{min-etal-2020-ambigqa}.
During experiments, we only consider test sets for inference.
Due to the small ClariQ test set (62 annotated queries) and the imbalanced feature, we use the entire ClariQ dataset, including train, validation, and test sets, for inference (299 annotated queries in total).
As for AmbigNQ, we use its validation set for inference since its test set is not publicly available.
Tab.~\ref{statistics} illustrates the statistics of the two datasets.

Note that we focus on a more general \ac{CNP} setting in conversational search, using only user queries without relying on any additional context for \ac{CNP}.
Hence, we do not consider the PACIFIC \cite{deng2022pacific} and Abg-CoQA \cite{guo2021abg} datasets, in which the ambiguity of a user query depends on a given document.
We do not use CLAQUA~\cite{xu2019asking}, which requires entities and and conversational history for \ac{CNP}. Additionally, we do not consider the datasets of MIMICS~\cite{zamani2020mimics} and MIMICS-Duo~\cite{tavakoli2022mimics}, which are not in the domain of conversational search and lacking clarification-need labels.

ClariQ\footnote{\url{https://github.com/aliannejadi/ClariQ}} is a mixed-initiative conversational search dataset with clarification-need labels annotated via crowdsourcing~\cite{aliannejadi2021building}.
The labels are ranged from 1 (no need for clarification) to 4 (clarification is necessary).
We follow the work from Meng et al.~\cite{meng2023system} to binarize the labels, with a label of 1 as ``not asking'' and labels of 2, 3, and 4 as ``asking'' a clarifying question. 

AmbigNQ\footnote{\url{https://huggingface.co/datasets/sewon/ambig_qa}} is an open-domain question answering dataset~\cite{min-etal-2020-ambigqa}.
The original one does not contain clarification-need labels, however, each question is accompanied by possible pairs of disambiguated questions and answers.
Arabzadeh
et al.~\cite{arabzadeh2022unsupervised} took advantage of the number of possible question-answer pairs to classify the level of ambiguity, and hence, we follow their work to collect pseudo-clarification-need labels. Specifically, if a question is linked to more than one question-answer pair, it is classified as ambiguous; otherwise, it is considered as unambiguous.

\begin{figure}[t]
     \centering
     \begin{tcolorbox}[notitle,boxrule=1pt,colback=gray!5,colframe=black, arc=2mm,width=\columnwidth]
\sf\textbf{Instruction}: Given the user query:\{query\}, predict if the query is vague or not. Please only return yes or no.
    \end{tcolorbox}
     \caption{
     Prompt for \acp{LLM} as clarification-need predictors.
     }
     \label{fig:baseline_prompt}
     \vspace{-2mm}
\end{figure}

\header{Baselines}
We focus on a zero-shot setting, and hence, we compare \our with baselines using \acp{LLM} as zero-shot clarification-need predictors. 
We use two open-source \acp{LLM}:
\begin{enumerate*}[label=(\roman*)]
\item Phi-3-mini-128k-instruct, and
\item Llama-3.1-8B-Instruct.
\end{enumerate*}
We use three commercial \acp{LLM}:
\begin{enumerate*}[label=(\roman*)]
\item GPT-3.5-turbo,
\item GPT-4, and
\item GPT-4o\-mini.
\end{enumerate*}
Fig.~\ref{fig:baseline_prompt} shows the prompt we use.
Note that we do not consider using \ac{CoT} to prompt \acp{LLM} directly for \ac{CNP}~\cite{deng2023prompting} because we focus on efficient \ac{CNP} in this work and using \ac{CoT} for \ac{CNP} significantly reduces inference efficiency.

We also consider an unsupervised \ac{CNP} model, MiniLm-ANC~\cite{arabzadeh2022unsupervised}. 
It first retrieves documents for a user query and then utilizes the coherency of documents retrieved for the user query: the more coherent the retrieved items are, the less ambiguous the query is, and the need for clarification decreases.

Moreover, we consider baselines accessing limited human-annotated clarifi\-cation-need labels.
We consider Llama-3.1-8B-Instruct, GPT-4, and GPT-4o-mini using the few-shot prompting, with 2, 4, or 6  demonstration examples, since they perform better in our preliminary experiments.
As mentioned before, we use the entire ClariQ for evaluation.
Therefore, we randomly sample demonstration examples from the AmbigNQ training set for few-shot prompting, inferring on ClariQ and AmbigNQ.

Note that we do not use \ac{CNP} baselines from \cite{deng2022pacific,deng2023prompting,guo2021abg} that focus on the \ac{CNP} scenarios where the ambiguity of a query depends on a given document.

\header{Evaluation metrics} 
As illustrated in Tab.~\ref{statistics}, our datasets show significant label imbalance.
As suggested by \cite{aliannejadi2021building}~\footnote{\url{https://github.com/aliannejadi/ClariQ}}, 
we adopt weighted Precision, Recall and F1.\footnote{Note that the score of weighted F1 may not fall between precision and recall. Please refer to~\url{https://scikit-learn.org/1.5/modules/generated/sklearn.metrics.f1_score.html}}
In the following sections, we will refer to them as Precision, Recall and F1 for simplification.
To evaluate \ac{CNP} efficiency, we measure \ac{CNP} inference latency averaged per query.

\header{Implementation details}
In \our, we use Llama-3.1-8B-Instruct or GPT-4o-mini for generating synthetic queries, considering the budget limitation.
We set the temperature of the \acp{LLM} to 1 to enhance the output diversity.
Overall, we generate 5,000 unique synthetic data, with 2,500 specific and ambiguous queries, respectively.
We use a BERT model (bert-base-uncased\footnote{\url{https://huggingface.co/google-bert/bert-base-uncased}}) as our efficient \ac{CNP} predictor and train it on the whole generated queries. 
We use 5e-5 as our learning rate, and a batch size of 64, with 3 training epochs.~\footnote{We use Hugging Face Trainer to implement our model fine-tuning. Other unmentioned hyperparameters are in the default settings.}
For baselines using \acp{LLM} as zero- and few-shot clarification-need predictors, we set the temperature of the \acp{LLM} to 0, making predicted results deterministic.
For MiniLM-ANC, we adhere to the setup shown in its original paper \cite{arabzadeh2022unsupervised}.

\section{Results and analysis}
\label{sec:results}

\begin{table}[t!]
\centering
\caption{
Results of \our and baselines using \acp{LLM} as zero/few-shot clarification-need predictors, on ClariQ and AmbigNQ, in terms of weighted Precision (P), Recall (R) and F1. 
\ac{CNP} inference latency is measured in seconds on a single GPU (A40), averaged per query.
The best value in each column is marked in \textbf{bold}, and the second best is \underline{underlined}.
} 
\resizebox{\textwidth}{!}{
\begin{tabular}{l cccccccccccc}
\toprule[1pt]
\multicolumn{1}{l}{{\multirow{2}{*}{Methods}}} & \multicolumn{4}{c}{ClariQ ($\%$)} & \multicolumn{4}{c}{AmbigNQ ($\%$)}
\\ \cline{2-9} 
\multicolumn{1}{c}{}& 
\multicolumn{1}{c}{P} & \multicolumn{1}{c}{R} & \multicolumn{1}{c}{F1} & \multicolumn{1}{c}{Latency} 
& \multicolumn{1}{c}{P} & \multicolumn{1}{c}{R} & \multicolumn{1}{c}{F1}& \multicolumn{1}{c}{Latency} \\

\hline

MiniLm-ANC~\cite{arabzadeh2022unsupervised} & 81.25  & 69.13 & 70.32 & 0.30  & 53.15  & 47.06 & 45.46 & 0.30 \\  

\midrule
Phi-3-mini-128k-instruct (0-shot)

&\textbf{89.17}	&13.38	&4.73	&0.21
&43.18	&41.41	&24.63 &0.22
\\ 
Llama-3.1-8B-Instruct (0-shot) 
&79.00	&66.56	&71.53 &0.08
&54.01 &55.19 & \underline{54.28} &0.08
 \\
GPT-3.5-turbo (0-shot)      
&79.49	&56.19	&63.66	&0.42
&54.25	&49.00	&47.52	&0.45
\\ 
GPT-4 (0-shot)        
&86.28	&29.77	&32.85	&0.58
&\underline{57.14}&42.31	&27.34	&0.65
\\ 
GPT-4o-mini (0-shot)   
&85.90	&56.86	&63.92	&0.51
&52.05	&42.96	&32.03	&0.47
\\\hline
Llama-3.1-8B-Instruct (2-shot)   
&78.72&75.25&76.88 &0.08
&55.42&\textbf{57.84}&54.19 &0.07

\\

Llama-3.1-8B-Instruct (4-shot)  
&79.20&67.56&72.26 &0.08
&52.79&53.25&52.98 &0.08
\\

Llama-3.1-8B-Instruct (6-shot)  
&79.52&62.88&68.98 &0.09
&51.78&51.65&51.71 &0.08

\\ GPT-4 (2-shot) &\underline{87.02} &50.17 &57.33 &2.37
&\textbf{59.28} &44.06 &32.20 &1.44
\\ 
GPT-4 (4-shot) &86.95 &49.50 &56.67 &0.52 
&52.96 &44.56 &37.36 &0.55
\\ 
GPT-4 (6-shot) &86.57 &57.53 &64.48 &1.45 
&49.49 &43.96  &39.74 &1.50
\\

GPT-4o-mini (2-shot) 
&86.78 &59.20& 65.97 &0.60
&53.68&43.16&31.70 &0.48
\\

GPT-4o-mini (4-shot)  
&86.62&72.58&76.99 &0.48
&55.36&44.26&34.52 &0.48
\\

GPT-4o-mini (6-shot)  
&85.85&73.24& 77.41 &0.48
&55.25&43.86&33.28  &0.51
 \\
 \hline

\our (Llama-3.1-8B-Instruct) specific $\rightarrow$ ambiguous 
&86.65 &\textbf{88.63} &\textbf{86.99}  &0.01
&54.98 &48.85 &46.68 &0.01
\\
\our (GPT-4o-mini) specific $\rightarrow$ ambiguous  
&82.40 &\underline{76.59} &\underline{79.02}  &0.01
&56.41 &\underline{55.94} &\textbf{56.13} &0.01
\\

\bottomrule[1pt]
\end{tabular}
}
\label{zeroshotvsours}
\end{table}

To answer \ref{RQ1}, Tab.~\ref{zeroshotvsours} shows \ac{CNP} quality and efficiency of \our, and baselines using \acp{LLM} as zero/few-shot clarification-need predictors, on ClariQ and AmbigNQ.
We have three main observations.

First, \our using GPT-4o-mini for the generation of synthetic queries outperforms all baselines in terms of F1 in both datasets.
Specifically, it exceeds the best baseline in ClariQ, GPT-4o-mini (6 shots), by 1.61 F1 points, and surpasses the best baseline in AmbigNQ, Llama-3.1-8B-Instruct (0 shots), by 1.85 F1 points.

Second, baselines demonstrate a large variation in \ac{CNP} quality across datasets.
For example,
GPT-4o-mini (6-shot) achieves an F1 score of 77.41 on ClariQ, but its performance drops significantly to 33.28 on AmbigNQ.
However, \our with GPT-4o-mini performs consistently well across both datasets, indicating it has better generalizability.

Third, \our is much more efficient than all baselines in terms of \ac{CNP} latency.
For example, Zef-CNP using GPT-4o-mini for synthetic query generation is approximately 80 times more efficient than the zero-shot baseline using the same LLM, on ClariQ. Moreover, it is around 90 times faster than the 2-shot baseline using GPT-4o-mini. However, we notice that there is inconsistency regarding the latency of OpenAI's API, which is also indicated by other research ~\cite{irugalbandara2023scaling}.


\subsection{Impact of the choice of \acp{LLM} on synthetic data generation} 
To answer \ref{RQ2}, Tab.~\ref{zeroshotvsours} also presents \ac{CNP} results of \our using open- and closed-source \acp{LLM} 
for synthetic data generation.
We find that \our with Llama-3.1-8B-Instruct performs much better than \our with GPT-4o-mini on ClariQ, but performs worse than many baselines on AmbigNQ.
In contrast, \our with GPT-4o-mini consistently shows advantages over baselines.
This finding indicates that different \acp{LLM} can excel in specific datasets, emphasizing the need to choose a versatile model like GPT-4o-mini for \our.
Moreover, combining synthetic query data generated from various \acp{LLM} might be able to further improve \our's performance, but we leave it to the future research.

\subsection{Analysis of \prompt and \seq}
\label{sec:ablation}

\begin{table}[t]
\centering
\caption{
Results of \our in different settings, on ClariQ and AmbigNQ, in terms of weighted Precision (P), Recall (R), and F1.
The best value in each column for each category is marked in \textbf{bold}. 
}
\resizebox{\textwidth}{!}{
\begin{tabular}{l ccc ccc}
\toprule[1pt]
\multicolumn{1}{c}{{\multirow{2}{*}{Methods}}} & \multicolumn{3}{c}{ClariQ ($\%$)} & \multicolumn{3}{c}{AmbigNQ ($\%$)}

\\ \cline{2-4}   \cline{5-7} 
\multicolumn{1}{c}{}
& \multicolumn{1}{c}{P} & \multicolumn{1}{c}{R} & \multicolumn{1}{c}{F1} 

& \multicolumn{1}{c}{P} & \multicolumn{1}{c}{R} & \multicolumn{1}{c}{F1} 
 
\\ 

\hline

\our (Llama-3.1-8B-Instruct) specific $\rightarrow$ ambiguous 
&\textbf{86.65} &\textbf{88.63} &\textbf{86.99}

&54.98 &48.85 &46.68 

\\
\our (Llama-3.1-8B-Instruct) ambiguous $\rightarrow$ specific &79.61 &84.62 &81.67 &\textbf{56.64}&\textbf{58.99} &52.88 
\\
w/o \seq &83.42 &87.63 &83.01 &54.69 &57.19 &\textbf{53.81} 
\\
w/o \seq, w/o \prompt 
&80.28 &84.62 &82.05  
 
&54.04 &53.55 &53.75 \\
\hline

\our (GPT-4o-mini) specific  $\rightarrow$ ambiguous 
&\textbf{82.40} &76.59 &79.02

&\textbf{56.41} &55.94 &\textbf{56.13} 

\\
\our (GPT-4o-mini) ambiguous $\rightarrow$ specific 
&81.32 &\textbf{85.28} &\textbf{82.83} 

&55.89 &57.94 &55.22 

\\
w/o \seq  &79.93 &79.93 &79.93 &55.86 &\textbf{58.34} &53.85 
\\
w/o \seq, w/o \prompt 
&81.06 &79.26 &80.11 

&55.83 &49.50 &47.37 \\

\bottomrule[1pt]
\end{tabular}
}

\label{tab:ablation}
\end{table}

To answer \ref{RQ3}, Tab.~\ref{tab:ablation} shows the \ac{CNP} quality of \our in four settings of synthetic query generation:
\begin{enumerate*}[label=(\roman*)]
\item generating a specific query first, followed by an ambiguous one in \seq with \prompt,
\item generating an ambiguous query first, followed by a specific one in \seq with \prompt,
\item generating both queries separately without \seq but with \prompt(Eq.~\ref{q1}), and
\item further removing \prompt.
\end{enumerate*}
We still use Llama-3.1-8B-Instruct and GPT-4o-mini for data generation.
We have two key findings.

First, the generation order of specific/ambiguous queries impacts \our's performance.
Different \acp{LLM} on different datasets prefer different generation orders.
For example, Llama-3.1-8B-Instruct performs better when generating a specific query first on ClariQ but prefers generating an ambiguous query first on AmbigNQ. 
In contrast, GPT-4o-mini shows the opposite trend.

Second, removing \seq, or further removing \prompt, results in poorer performance in most cases. 
For example, for ClariQ, when using Llama-3.1-8B-Instruct, removing \seq results in a significant drop of 3.98 F1 points, compared to generating specific queries first with both \prompt and \seq; further removing \prompt leads to an additional drop of 0.96 F1 points. 
It indicates the effectiveness of \prompt and \seq in improving the synthetic specific/ambiguous query generation in a zero-shot manner.

\begin{figure}[t!]
    \centering

    \begin{subfigure}{0.44\linewidth}
      \includegraphics[width=\linewidth]{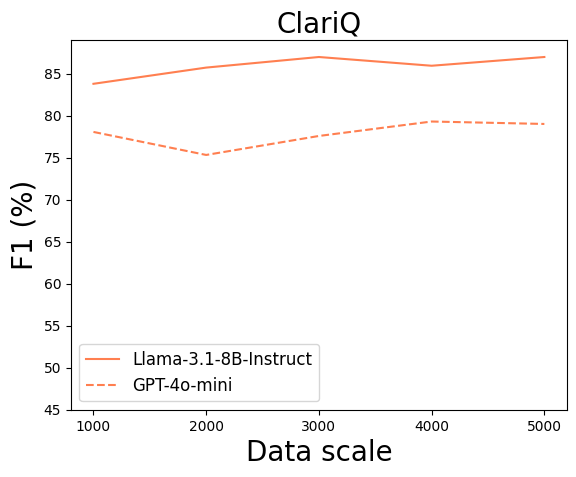}
    \end{subfigure}
    ~~
    \begin{subfigure}{0.44\linewidth}
      \includegraphics[width=\linewidth]{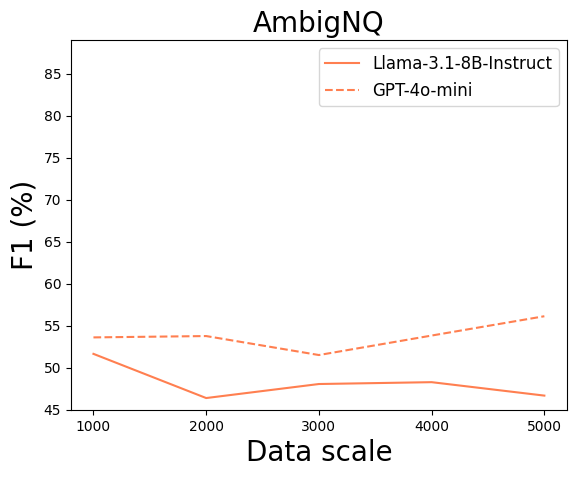}
    \end{subfigure}
    \caption{\ac{CNP} results of \our in terms of weighted F1 w.r.t. the impact of different scales of generated data by Llama-3.1-8B-Instruct and GPT-4o-mini, on ClariQ and AmbigNQ.
    } 
    \label{datascale}
\end{figure}

\subsection{Impact of synthetic data scale on \ac{CNP} quality} 
To answer \ref{RQ4}, Fig.~\ref{datascale} illustrates \ac{CNP} quality on ClariQ and AmbigNQ, in terms of F1 w.r.t.\ the impact of different scales of synthetic data generated by Llama-3.1-8B-Instruct and GPT-4o-mini, and each data scale consists of an equal split between specific and ambiguous queries. 
We observe that \ac{CNP} quality improves when our efficient \ac{CNP} model (BERT) is trained with larger amounts of synthetic data, despite some fluctuations, except when using BERT fine-tuned on synthetic queries generated by Llama-3.1-8B-Instruct, and inferring on AmbigNQ.
Surprisingly, only using 1000 synthetic queries already results in a decent performance in some cases.
For example, training in 1,000 queries generated by Llama-3.1-8B-Instruct achieves an F1 score of 83.80 in ClariQ, which already surpasses all baselines listed in Tab.~\ref{zeroshotvsours}.

\subsection{Room for improvement}

To show the room for improvement of \our, 
Tab.~\ref{finetuned} presents the \ac{CNP} results of \acp{PLM} fine-tuned on the training set of AmbigNQ (8,028 queries) and our 5,000 synthetic queries with clarification-need labels (referred to as \data), inferring on AmbigNQ.
We observe that training on large-scale in-domain data still results in better \ac{CNP} quality than using our synthetic data.
It suggests that there is a space for the improvement of our work.

\begin{table}[t!]
\centering
\caption{
\ac{CNP} results of \acp{PLM} fine-tuned on human-annotated clarification-need labels (the training set of AmbigNQ with 8,028 examples) and our synthetic clarification-need labels (referred to as \data) on AmbigNQ. The best value in each column is marked in \textbf{bold}.
}
\resizebox{\textwidth}{!}{
\begin{tabular}{llccc}
\toprule[1pt]
\multicolumn{1}{l}{{\multirow{2}{*}{Methods}}} 
&\multicolumn{1}{l}{{\multirow{2}{*}{Training source}}} 

& \multicolumn{3}{c}{AmbigNQ ($\%$)}

\\ \cline{3-5}  
\multicolumn{1}{c}{}

&\multicolumn{1}{c}{}& \multicolumn{1}{c}{P} & \multicolumn{1}{c}{R} & \multicolumn{1}{c}{F1} 
 
\\ 
\hline
roberta-base    
&AmbigNQ training set&\textbf{67.24} & \textbf{66.68} & \textbf{66.86}

\\ 
bert-base-uncased    
&AmbigNQ training set&66.21 &65.33 &65.56	

\\ 
\hline

\our (Llama-3.1-8B-Instruct) specific $\rightarrow$ ambiguous
&\data &54.98 &48.85 &46.68 

\\
\our (GPT-4o-mini) specific $\rightarrow$ ambiguous
&\data & 56.41 &55.94 &56.13

\\

\bottomrule[1pt]
\end{tabular}
}

\label{finetuned}
\end{table}

\section{Conclusions}
\label{sec:conclusions}

In this paper, we have proposed \our, a zero-shot and efficient \ac{CNP} framework for mixed-initiative conversational search, which uses \acp{LLM} to generate synthetic \ac{CNP} training data in a zero-shot way and then trains efficient \ac{CNP} predictors on the synthetic data.
\our eliminates the need for human-annotated clarification-need labels during training and avoids relying on \acp{LLM} or retrieved documents at query time.
To improve the quality of zero-shot generation of specific and ambiguous queries, we devise \prompt and \seq. 
\prompt guides \acp{LLM} to generate topics and information needs, step by step, before generating synthetic queries.
\seq enhances \prompt by enabling counterfactual generation of specific and ambiguous queries.
Experimental results show that \our (fully zero-shot) outperforms baselines using \acp{LLM} as zero/few-shot clarification-need predictors, in terms of both \ac{CNP} effectiveness and efficiency, on two benchmark datasets, ClariQ and AmbigNQ.

\header{Broader impact}
We believe that our method has two broader impacts.
First, \our offers a solution for domain-specific scenarios where no human-annotated \ac{CNP} training data are available.
Second, \our operates without using \acp{LLM} or retrieval at query time, significantly reducing latency and improving \ac{CNP}'s real-world applicability in practice.

\header{Limitations and Future work}
Our work has three limitations.
First, as shown in Tab. \ref{finetuned}, there is room to further improve the quality of synthetic queries.
One factor that may limit the quality of synthetic data is that we fine-tune efficient \ac{CNP} models using synthetic data generated by a single \ac{LLM}.
Only relying on one \ac{LLM} might introduce biases in the data. To mitigate it, we plan to fine-tune \ac{CNP} models on data generated by multiple \acp{LLM}. 
Second, we only use BERT as our efficient \ac{CNP} model. 
Although BERT is widely used, it is also worthwhile to investigate the results of simpler models fine-tuned on our synthetic data. 
Third, we only focus on English datasets and it is interesting to explore the suitability and robustness of our proposed method on multi-lingual aspect~\cite{sun2021conversations}. 

\header{Acknowledgement}
We thank our reviewers for their feedback.
Our work is supported by the Swiss National Science Foundation (SNSF), under the project PACINO (Personality And Conversational INformatiOn Access).

%
%
%

\bibliographystyle{splncs04}
\bibliography{references}

\end{document}